\documentclass[english]{article}
\usepackage[T1]{fontenc}
\usepackage[latin9]{inputenc}
\usepackage{geometry}
\usepackage{color,times}
\usepackage{graphicx}
\usepackage{babel}
\usepackage{epstopdf}
\usepackage{epsfig,cite}

\date{}
\begin{document}

\title{Kak's three-stage protocol of secure quantum communication revisited:
Hitherto unknown strengths and weaknesses of the protocol}
\author{Kishore Thapliyal\footnote{Email: tkishore36@yahoo.com} \, and Anirban Pathak\footnote{Email: anirban.pathak@gmail.com} \\
Jaypee Institute of Information Technology, A-10, Sector 62. Noida,
UP-201307}
\maketitle

\begin{abstract}
Kak's three-stage protocol for quantum key distribution is revisited with special focus
on its hitherto unknown strengths and weaknesses. It is shown that this protocol can be used for secure direct quantum communication. Further, the implementability of this protocol in the realistic situation is analyzed by considering various Markovian noise models. It is found that the Kak's protocol and its variants in their original form can be implemented only in a restricted class of noisy channels, where the protocols can be transformed to corresponding protocols based on logical qubits in decoherence free subspace. Specifically, it is observed that Kak's protocol can be implemented in the presence of  collective rotation and collective dephasing noise, but cannot be implemented in its original form in the presence of other types of noise, like amplitude damping and phase damping noise. Further, the performance of the protocol in the noisy environment is quantified by computing average fidelity under various noise models, and subsequently a set of preferred states for secure communication in noisy environment have also been identified.
\end{abstract}

\section{Introduction}
In 1984, Bennett and Brassard proposed the first protocol for quantum key distribution (QKD) \cite{bennett1984quantum}. It succeeded to draw the attention of the cryptography community immediately as it could provide unconditional security, which is a desired but unachievable feat in the classical world. Because of these interesting features of QKD, the pioneering work of Bennett and Brassard was followed by a  large number of protocols for QKD \cite{ekert1991quantum,bennett1992quantum,bennett1992B92,goldenberg1995quantum} and secure direct quantum communication \cite{bostrom2002deterministic,degiovanni2004quantum,lucamarini2005secure} (where prior generation of a key is not required) (see \cite{pathak2018quantum} for a review). Among these schemes,  only a few schemes have been realized experimentally (\cite{bennett1992experimental,zhao2006experimental,schmitt2007experimental,lo2014secure,hu2016experimental,zhang2017quantum} and references therein). Further, almost all the experimentally realized schemes of secure quantum communication are protocols for QKD. Only, recently a few schemes of secure direct quantum communication have been realized experimentally \cite{hu2016experimental,zhang2017quantum,cao2017direct}. This fact motivated us to look for simple schemes of secure direct quantum communication that can be realized experimentally.  During our investigation, we realized that there exists an experimentally implemented scheme for secure quantum communication,  which can be viewed as a scheme for secure direct quantum communication, but in the original proposal as well as in the follow-up works, it has been described as a scheme for QKD. 
Specifically, a three-stage protocol for QKD was proposed
by Kak in 2006 \cite{kak2006three} and experimentally implemented in 2013 by Mandal et al. \cite{mandal2013multi}. This scheme has certain advantages over the conventional BB84 protocol and its variants. For example, it does not require single photon source and can be implemented using multi-photon pulses \cite{mandal2013multi, chan2015security}. Further,  it can be modified to obtain three-stage quantum protocols for other  quantum communication
tasks.  For example, three-stage schemes
for quantum oblivious transform \cite{parakh2013quantum} has been proposed\footnote{Note that \cite{parakh2013quantum} contradicts the well established results of Ref. \cite{lo1997insecurity} and the protocol reported \cite{parakh2013quantum} is not loophole free.} using Kak's protocol. A quantum signature scheme \cite{kang2015quantum} and a public key cryptography scheme \cite{nikolopoulos2008applications}  based on Kak's three-stage protocol were also proposed. This protocol is also found useful in quantum handshake \cite{el2014ieee}, intensity-aware \cite{kak2012iaqc} and threshold quantum cryptography \cite{kak2013threshold}, and in a variant of it  where each pulse transmits more than one bit \cite{el2013implementation}. Attempts have also been made to reduce the number of rounds of quantum communication in the three-stage protocol by proposing single-stage and braided single-stage protocols \cite{darunkar2014braided}, where Bob is already aware of the unitary operation Alice has applied. However, this 1 stage variant of Kak's protocol fails to qualify as a scheme for QKD as it requires a prior knowlge of the unitary operation which equivalent to a pre-shared key.

All the above mentioned three-stage schemes are interesting in their own merit. However, the effects of noise on those schemes
are not rigorously studied. Of course, in Refs. \cite{wu2015three,parakh2016correcting} and Ref. \cite{chitikela2012noise}, it is claimed that effect of collective rotation (CR) noise and uniform distribution of error caused due to  different sources of noise on Kak's protocol have been studied, respectively, but these efforts were not mathematically rigorous.  Keeping this fact in mind, in the present
paper, the effects of different types of noise models (e.g., amplitude
damping (AD), phase damping (PD), collective dephasing (CD), CR)
on the Kak-type three-stage protocols of secure quantum communication
have been studied. Here, we have considered that the noise parameters remain same for each round of travel through the quantum channel for all three-stages of the particular quantum
communication process. In what follows, the effect of noise is illustrated by plotting
the fidelity of the expected quantum state and the produced quantum
state vs decoherence and other relevant parameters. In most of
the cases, we have observed that the effect of PD noise
is more than that of the AD noise for the same decoherence
rate. Very interestingly, it has been observed that Kak's protocol
only works under CR noise. It fails under AD, PD, and CD noise. This is so
because the Kraus operators of the noise models (except that of CR) do not commute with unitary operators used by Alice and
Bob in Kak's three-stage protocol. A similar conclusion holds for other
protocols of secure quantum communications that are based on Kak's
protocol. Finally, we have tried to propose some methods that may be adapted to circumvent this problem and implement Kak-type three-stage protocols in the presence of noise. 

The rest of the paper is organized as follows. In Sec. \ref{sec:review}, we briefly discuss the Kak's three-stage protocol and its origin of security and uniqueness in Section \ref{sec:review-sec}. Thereafter, the effect of noise on three-stage protocol is studied in the next two sections before concluding in Section \ref{sec:con}. 

\section{Kak's three-stage protocol \label{sec:review}}

To begin with, we briefly describe Kak's original three-stage protocol
for QKD \cite{kak2006three} which may be summarized in the following steps:
\begin{enumerate}
\item Alice prepares a single qubit quantum state $|\psi\rangle\, \in \left\{\left(\alpha |0\rangle+\beta |1\rangle\right),\left(\beta |0\rangle-\alpha |1\rangle\right) \right\}$. The basis
and the corresponding bit values for both orthogonal states has been
priorly decided, i.e., for sending a 0 (1) she prepares $\alpha |0\rangle+\beta |1\rangle$ $\left(\beta |0\rangle-\alpha |1\rangle\right)$.
\item Alice applies a unitary operator $U_{A}\equiv R\left(\theta\right)$
to transform the state $|\psi\rangle$ to $|\psi^{\prime}\rangle=R(\theta)|\psi\rangle$
and sends the transformed qubit $|\psi^{\prime}\rangle$ to Bob. Here,
the unitary operator used is a rotation operator $R(\theta)=\left[\begin{array}{cc}
\cos\theta & -\sin\theta\\
\sin\theta & \cos\theta
\end{array}\right]$.
\item Bob independently applies a unitary transformation $U_{B}\equiv R\left(\phi\right)$
to transform $|\psi^{\prime}\rangle$ into $|\psi^{\prime\prime}\rangle=R(\phi)|\psi^{\prime}\rangle=R(\phi)R(\theta)|\psi\rangle=R(\theta)R(\phi)|\psi\rangle$
(in the last step we have used the fact that two arbitrary rotation operators commute)
and sends it back to Alice.
\item This time Alice applies $U_{A}^{\dagger}=U_{A}^{-1}$ to transform
$|\psi^{\prime\prime}\rangle$ to $|\psi^{\prime\prime\prime}\rangle=R(\phi)|\psi\rangle$ and
sends the qubit again to Bob.
\item Bob applies $U_{B}^{\dagger}=U_{B}^{-1}$ to obtain $|\psi\rangle$
the state (bit value) Alice wanted to share.
\end{enumerate}
Although Kak introduced the above protocol as a protocol for QKD, it
is easy to recognize that Alice is not bound to send a random bit
value using this scheme. She can always send a sequence of meaningful
bits using this scheme and thus Kak's protocol should be viewed as
a protocol of quantum secure direct communication, where a message
can be transmitted directly without constructing a prior key. Once
we recognize this protocol as a scheme of quantum secure direct communication we can naturally extend
it to construct several other schemes of secure quantum communication
that are variants of direct communication (for a detail discussion see \cite{pathak2013elements} and references therein). There exist several schemes
for secure direct quantum communication \cite{bostrom2002deterministic,lucamarini2005secure,shukla2013improved,long2007quantum,hu2016experimental}. In fact, it is easy to show that famous BB84 scheme
can be transformed to a scheme for secure direct quantum communication if one allows Bob to use quantum
memory. To be precise, if Bob stores the string of single photons corresponding to message and checking qubits prepared randomly in $\left\{|0\rangle, |1\rangle \right\}$ and $\left\{|+\rangle, |-\rangle \right\}$ in a quantum memory until Alice discloses the positions and basis of the checking qubits. Using the same basis Bob performs a measurement of these verification qubits and announces the measurement outcomes, which help Alice in deciding whether to proceed with announcing the basis used to prepare the message qubits when error rate is below the threshold limit. Thus, Alice and Bob can perform a direct communication with no need of key generation. It may be noted that most of
the well known protocols for direct secure quantum communication schemes use quantum memory. For example,
we may briefly describe ping-pong protocol \cite{bostrom2002deterministic} for direct secure quantum communication as a scheme where
Bob prepares a Bell state and sends a qubit to Alice to encode her message keeping another qubit in a quantum memory before measuring both the qubits in Bell basis to extract the secret \footnote{An alternative definition exists in the literature \cite{long2007quantum} according to which Kak's scheme should be viewed as a scheme for deterministic QKD since it does not involve block transmission. However, a deterministic QKD scheme can be adapted to perform direct secure quantum communication if the sender encrypts the message with a randomly chosen private key before sending it to the receiver using deterministic QKD and revealing the key only when she ensures the secure transmission of ciphertext.}. Similarly, LM05 protocol \cite{lucamarini2005secure},
Shukla et al.'s protocol \cite{shukla2013improved}, recent experimental implementation by Zhang et al. \cite{zhang2017quantum} for direct secure quantum communication do use quantum memory. This is the point where the actual
strength of Kak's protocol lies. It does not require quantum memory.
 This is important as quantum memory is a very costly
resource and so far we do not have any good solution for a reliable quantum
register that can store qubits for a reasonable amount of time. To
the best of our knowledge, there exists only one proposal for direct secure quantum communication 
without quantum memory \cite{yang2014quantum}. In the Yang's scheme \cite{yang2014quantum}, LM05 protocol \cite{lucamarini2005secure} of
direct secure quantum communication was suitably modified to obtain a scheme for QSDC without quantum
memory. 

Further, several direct communication schemes have been modified to obtain solutions of various cryptographic tasks, such as controlled \cite{pathak2015efficient,thapliyal2015applications}, asymmetric \cite{banerjee2017asymmetric} and multiparty \cite{banerjee2017quantum} variants of direct communication schemes, quantum e-commerce \cite{shukla2017semi}, quantum voting \cite{thapliyal2017protocols}, quantum sealed-bid auction \cite{sharma2017quantumauction}, quantum private comparison \cite{thapliyal2016orthogonal,shukla2017semi}. Therefore, the use of quantum memory plays an instrumental role in the implementation of some of these schemes as well and modified Kak's protocol can help us to circumvent the use of quantum memory in the experimental realization of the above mentioned cryptographic tasks.

\section{Nature and origin of security \label{sec:review-sec}}

Schemes for secure quantum communication can be broadly divided into
two types, orthogonal-state-based schemes and conjugate-coding-based
scheme. Orthogonal state based schemes, such as Goldenberg-Vaidman protocol \cite{goldenberg1995quantum}, use the same basis for encoding,
decoding and eavesdropping checking. Whereas in the conjugate-coding-based
schemes, like BB84 protocol \cite{bennett1984quantum}, the security comes from non-commutativity and no-cloning theorems. The origin of unconditional security in the quantum domain can also be understood from the splitting of information. Precisely, the sender splits useful information in multiple quantum and classical pieces and ensures all of them remain unavailable to unintended intruders until the secure communication is accomplished. In case of Kak's protocol, the sender prepares a quantum piece (qubit) and withholds a classical information (unitary $U_{A}$) until the receiver also composes a classical piece (another unitary $U_{B}$) to perform the cryptographic task. As $U_{A}$ and $U_{B}$ are not directly used for the encoding, decoding or eavesdropping checking, and as these operations can be done using orthogonal states, this protocol can be implemented as an orthogonal-state-based protocol.

\section{Effect of noise \label{sec:noise}}

The beauty of the Kak's protocol lies in the fact that $U_{A}$ and $U_{B}$
commute, and the original security proof is restricted to the ideal
situation, where there is no noise present in the channel between Alice
and Bob. However, in any practical implementation of the scheme, it would be impossible to completely circumvent noise. Keeping this in mind, in what follows, we wish to investigate the effect of various
types of noise on  Kak's protocol and its variants. The effect of noise on Kak's
protocol can be studied using Kraus operators for various noise models.

Mathematically, evolution of a single qubit quantum state $\rho$ in the noisy environment can be described using the Kraus operator formalism
as \cite{breuer2002theory,nielsen2010quantum,preskill1998lecture,thapliyal2015quasiprobability}
\begin{equation}
\rho_{k}=\sum_{i}E_{i}^{k}\rho\left(E_{i}^{k}\right)^{\dagger},\label{eq:noise-effected-density-matrix-1}
\end{equation}
where $E_{i}^{k}$s are the Kraus operator for a specific noise model (displayed as superscript $k$) under consideration.
Before we proceed further, we need to state the Kraus operators for various noise models. In the following subsection, we have listed the Kraus operators for various noise models that are investigated in this work. 

\subsection{Kraus operators for various noise models}
\begin{enumerate}
\item AD noise: The spontaneous emission from a high energy state is modeled
by the following set of Kraus operators \cite{nielsen2010quantum,preskill1998lecture,thapliyal2015quasiprobability}:
\begin{equation}
E_{0}^{A}=\left[\begin{array}{cc}
1 & 0\\
0 & \sqrt{1-\eta_{A}}
\end{array}\right],\,\,\,\,\,\,\,\,\,\,\,\,\,\,\,E_{1}^{A}=\left[\begin{array}{cc}
0 & \sqrt{\eta_{A}}\\
0 & 0
\end{array}\right],\label{eq:Krauss-amp-damping}
\end{equation}
where the decoherence rate $\eta_{A}$ such that $0\leq\eta_{A}\leq1$ depends on the interaction between the system and the environment.
\item PD noise: This dephasing noise model is
described by the Kraus operators \cite{nielsen2010quantum,preskill1998lecture}:
\begin{equation}
E_{0}^{P}=\left[\begin{array}{cc}
1 & 0\\
0 & \sqrt{1-\eta_{P}}
\end{array}\right],\,\,\,\,\,\,\,\,\,\,\,\,\,\,\,E_{1}^{P}=\left[\begin{array}{cc}
0 & 0\\
0 & \sqrt{\eta_{P}}
\end{array}\right],\label{eq:Krauss-phase-damping}
\end{equation}
with the decoherence rate
$\eta_{P}$ ($0\leq\eta_{P}\leq1)$ involves interaction without energy loss.
\item CD noise model: A coherent effect of environment on all the qubits traveling through a channel is studied as collective noise models \cite{zanardi1997noiseless,bourennane2004decoherence}. This kind of noise model is described by unitary operations, unlike Kraus operators of AD or PD channels. Specifically, collective noise is studied as CD and CR noise models. The effect of CD noise
is characterized (\cite{sharma2016verification} and references therein) by $E^{D}\left|0\right\rangle =\left|0\right\rangle$ and $E^{D}\left|1\right\rangle =\exp\left(i\Phi\right)\left|1\right\rangle.$
One can easily obtain that $E^{D}=\left[\begin{array}{cc}
1 & 0\\
0 & \exp\left(i\Phi\right)
\end{array}\right]$ is a phase gate only. Here, $\Phi$ is the noise parameter that remains the same for all the travel qubits at any instant of time. However, it can take different values while independent use of a channel at different times.

\item CR noise model: Similar to the CD noise model, this type of noise (\cite{sharma2016verification} and references therein)  is defined to affect as $E^{R}\left|0\right\rangle =\cos\Theta\left|0\right\rangle +\sin\Theta\left|1\right\rangle$ and $E^{R}\left|1\right\rangle =-\sin\Theta\left|0\right\rangle +\cos\Theta\left|1\right\rangle ,$
which can be easily defined due to the application of a unitary rotation $E^{R}=\left[\begin{array}{cc}
\cos\Theta & -\sin\Theta\\
\sin\Theta & \cos\Theta
\end{array}\right].$ Quite similar to the CD noise, here $\Theta$ is the noise parameter that may change with time and affects all the
travel qubits in the same way. 
\end{enumerate}

\subsection{Commutativity of the rotation operator used in Kak's protocol and
the Kraus operators}

Using the open quantum system approach mentioned in Eq. (\ref{eq:noise-effected-density-matrix-1}), we can summarize the evolution of a single qubit quantum state under the Kak's three-stage protocol in the noisy environment as

\begin{equation}
\rho_{k}=\sum_{i,j,l}\left(\left(R\left(\phi\right)\right)^{\dagger}E_{i}^{k}R\left(\left(\theta\right)\right)^{\dagger}E_{j}^{k}R\left(\phi\right)E_{l}^{k}R\left(\theta\right)\right)\rho\left(\left(R\left(\phi\right)\right)^{\dagger}E_{i}^{k}R\left(\left(\theta\right)\right)^{\dagger}E_{j}^{k}R\left(\phi\right)E_{l}^{k}R\left(\theta\right)\right)^{\dagger}.\label{eq:Kak-noise}
\end{equation}
Here, $k\in\left\{\mathrm{AD},\mathrm{PD}\right\}$ corresponds to the type of noise model under consideration, and $\rho=|\psi\rangle\langle\psi|$ is the single qubit initial state prepared by Alice. The single qubit state is rotated by an angle $\theta\,\left(\phi\right)$ in the Bloch sphere by Alice's (Bob's) operation in Step 1 (3). Additionally, different $i,\,j,$ and $l$ in the subscript represent independent effects of noise during Alice--to--Bob, Bob--to--Alice, and Alice--to--Bob travels of single qubit, respectively.

At a first glance, one can easily conclude that the beauty of the Kak's protocol (i.e., commutativity of rotation operators by Alice and Bob) could only be preserved if the rotation operators commute with the Kraus operators for various noise models.

To begin with, let us consider a simple situation in which Kak's protocol
is implemented using an AD channel, and in the first two stages of the
protocol (i.e., from Alice--to--Bob and Bob--to--Alice journey), noise
affects the qubit via $E_{0}^{A}$, in this situation, instead of
$|\psi^{\prime\prime}\rangle=R(\phi)R(\theta)|\psi\rangle=R(\theta)R(\phi)|\psi\rangle$,
Alice would receive $|\psi^{\prime\prime}\rangle=E_{0}^{A}R(\phi)E_{0}^{A}R(\theta)|\psi\rangle$.
Note that Alice would be able to remove her encryption $U_{A}=R(\theta)$
by applying $U_{A}^{\dagger}=U_{A}^{-1}$ if and only if $R(\theta)$
commutes with $E_{0}^{A}$ (i.e., iff $\left[E_{0}^{A},R(\theta)\right]=0$).
However, we can easily compute that 
\begin{equation}
\left[E_{0}^{A},R(\theta)\right]=E_{0}^{A}R(\theta)-R(\theta)E_{0}^{A}=\left(1-\sqrt{1-\eta}\right)\sin\theta\left[\begin{array}{cc}
0 & 1\\
1 & 0
\end{array}\right].\label{eq:amp0-r}
\end{equation}
For this commutator to vanish, i.e., $\left[E_{0}^{A},R(\theta)\right]=0$, we require either $\eta=0$ or $\theta=0$.
The former case corresponds to noiseless situation while the latter
case corresponds to no rotation applied by Alice in the Bloch sphere in Kak's protocol, i.e., $R(\theta)$
becomes identity and eavesdropper's ignorance becomes zero. These
are trivial cases, and the analysis shows that in the above situation
Kak's protocol does not work in its original form. The observation
can be further strengthened by noting that 
\begin{equation}
\left[E_{1}^{A},R(\theta)\right]=E_{1}^{A}R(\theta)-R(\theta)E_{1}^{A}=-\sqrt{\eta}\sin\theta\left[\begin{array}{cc}
1 & 0\\
0 & -1
\end{array}\right],\label{eq:amp1-r}
\end{equation}
where the results obtained from Eq. (\ref{eq:amp0-r}) remains valid.
Hence, it can be summarized that Kak's three-stage protocol fails
under AD noise as the rotation operator of Alice operated
in the second step of Kak's protocol will not be nullified by the
inverse operator of the same rotation operator applied in the fourth
step of the protocol. 

Similar studies can be performed for other kind of noises, like PD
noise. In fact, it is observed that both $E_{0}^{P}$ and $E_{1}^{P}$ result in the same conclusion as we obtained for the AD
noise channel. Similar investigations  over the CD and CR noise is quite easy to perform as the effect of noise is characterized using unitary operators in both these cases. Specifically, the CD noise leads to the same result that the rotation operator commutes only in the ideal condition, and in the noisy scenario, for $\Phi=2n\pi$ with an integer $n$ (as the unitary for CD noise reduces to an identity)\footnote{Here, it may be noted that although Kak's protocol in its original form would not work under the CD noise, there are techniques to use logical qubits and thus to exploit the advantage of a decoherence free subspace to realize Kak's protocol in presence of CD noise \cite{li2008efficient,sharma2016verification}, but no such decoherence free subspace is known for the AD and PD noise.}. It would also be worth noting
here that CR noise does not affect the protocol as
two arbitrary rotation operators always commute with each other. Hence, Kak's
protocol would work under CR noise. One such attempt to analyze the Kak's three-stage protocol over CR noise in multi-photon case \cite{wu2015three}, and it is shown to possess higher error rate tolerance than the single photon case. 

It is expected
that a similar study on the squeezed generalized amplitude damping
channel would also lead to the same conclusion as generalized
amplitude damping and AD noise channels are only the limiting
cases of squeezed generalized amplitude damping channel. 

\section{Formal investigation on the effect of noise on the Kak's three-stage
protocol \label{sec:noise-kak}}

The effect of noise can be formally investigated by comparing
the quantum state $\rho_{k}$ produced in the noisy environment with
the state $\rho=|\psi\rangle\langle\psi|$ which was expected at the Bob's port after three-stages of quantum communication in the absence
of noise. The comparison can be performed using fidelity 
\begin{equation}
F=\langle\psi|\rho_{k}|\psi\rangle,\label{eq:fidelity}
\end{equation}
which is the square of the conventional definition of fidelity. 
In addition, for the convenience of discussion, an arbitrary single qubit quantum
state which is to be transmitted by Alice in Step 1 (before application
of $U_{A}$) to send a bit value "0" can be assumed as $|\psi_1\rangle=\cos\xi|0\rangle+\sin\xi|1\rangle$; whereas Alice has to send an orthogonal state $|\psi_2\rangle=\sin\xi|0\rangle-\cos\xi|1\rangle$ to send a bit value of "1".
Therefore, the initial density matrix will be 
$\rho=|\psi\rangle\langle\psi|$ with $|\psi\rangle\in\left\{|\psi_1\rangle,|\psi_2\rangle\right\}$.

In the presence of AD noise (i.e., when the qubit is subjected to AD noise), using Eqs. (\ref{eq:Krauss-amp-damping}) and (\ref{eq:fidelity}) a closed form analytic expression of fidelity is computed as 
\begin{equation}
\begin{array}{lcl}
F_{AD} & = & \frac{1}{16}\left[-\eta\left(\eta^{2}-3\left(\sqrt{1-\eta}+2\right)\eta+7\sqrt{1-\eta}+9\right)+4\left(\sqrt{1-\eta}+3\right)\right.\\
 & - & \left.(\eta-1)\left(\eta\left(\eta+3\sqrt{1-\eta}-5\right)-4\sqrt{1-\eta}+4\right)\cos(4\xi)\right],
\end{array}\label{eq:fad}
\end{equation}
which is averaged over two possible choices of the initial state by Alice, depending up on the bit value of the message she wishes to send.

Along the same line, a similar study over purely dephasing kind of noise (i.e., PD noise)  using Eqs. (\ref{eq:Krauss-phase-damping}) and (\ref{eq:fidelity}) led to the following compact expression 
\begin{equation}
F_{PD}=\frac{1}{8} \left(\left(-\sqrt{1-\eta } \eta +3 \eta +4 \sqrt{1-\eta }-4\right) \sin ^2(2 \xi)-3 \eta +8\right).
\label{eq:fpd}
\end{equation}

The fidelity of the quantum state received by Bob over collective noisy channels with that of Alice's initially prepared state is computed as
\begin{equation}
F_{CD}=\frac{1}{32} \left(6 \cos ^2(2 \theta ) \cos (2 \Phi )+\sin ^2(2 \theta ) (15 \cos (\Phi )+\cos (3 \Phi ))+5 \cos (4 \theta )+21\right)
\label{eq:fcd}
\end{equation}
and 
\begin{equation}
F_{CR}=\cos ^2(3 \Theta),
\label{eq:fcr}
\end{equation}
for CD and CR channels, respectively.

Note that the choice of state parameter $\xi$ by Alice is a public  knowledge (i.e., decided at the beginning of the protocol by Alice and Bob), which is a continuous variable in the domain $\left\{0,2\pi\right\}.$ Therefore, before reaching to any conclusion from Eqs. (\ref{eq:fad}) and (\ref{eq:fpd}), it would also be imperative to compute the average fidelity by taking into account all possible choices of $\xi$ using 
\begin{equation}
F_{k}^{\mathrm{av}}=
\frac{1}{2\pi}{\int_{0}^{2\pi} F_{k} d\theta}.\label{eq:avf}
\end{equation}
The average fidelity expressions calculated over AD and PD noise channels are
\begin{equation}
F_{AD}^{\mathrm{av}}=\frac{1}{16} \left(4 \left(\sqrt{1-\eta }+3\right)-\eta  \left(\eta ^2-3 \left(\sqrt{1-\eta }+2\right) \eta +7 \sqrt{1-\eta }+9\right)\right),\label{eq:avfad}
\end{equation}
and 
\begin{equation}
F_{PD}^{\mathrm{av}}=\frac{1}{16} \left(\sqrt{1-\eta }+3\right) (4-\eta),
\label{eq:avfpd}
\end{equation}
respectively.

Similarly, average fidelity in the case of qubit subjected to CR noise is
\begin{equation}
F_{CD}^{\mathrm{av}}=\frac{1}{64} (15 \cos (\phi )+6 \cos (2 \phi )+\cos (3 \phi )+42).
\label{eq:avfcd}
\end{equation}
Also, from Eq. (\ref{eq:fcr}) one can conclude that $F_{CR}^{\mathrm{av}}=F_{CR}$ as the obtained expression for $F_{CR}$ is independent of the choice of the initial state.

\begin{figure}
\includegraphics[scale=0.58]{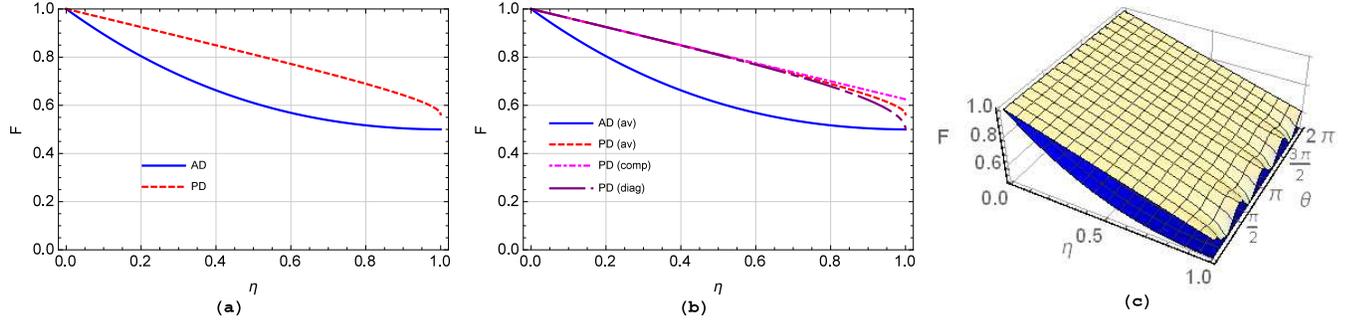}
\caption{\label{fig:Kak1}(Color online) Variation of average fidelity as a function of
noise parameters of AD and PD noise is shown in (a). In (b), fidelity for the choice of initial states in the computational $\left\{|0\rangle,|1\rangle\right\}$ and diagonal
$\left\{|+\rangle,|-\rangle\right\}$ basis is also shown. In (c), dependence on the choice of initial state is illustrated through a three-dimensional plot where light yellow (dark blue) colored surfaces correspond to fidelity calculated over AD and PD noise models, respectively.}
\end{figure}

Finally, we have established dependence of fidelity (in Eqs. (\ref{eq:fad}) and (\ref{eq:fpd})) on the state parameter, and also shown its variation along with average fidelity (in Eqs. (\ref{eq:avfad}) and (\ref{eq:avfpd})) in Figure \ref{fig:Kak1}. 
The obtained fidelity and average fidelity over a PD noise channel is always better than that for AD channel (cf. Figure \ref{fig:Kak1} (a) and (c)). Further, in Figure \ref{fig:Kak1} (b), we compare the obtained average fidelity with that of the fidelity for the initial states chosen from the computational $\left\{|0\rangle,|1\rangle\right\}$ and diagonal
$\left\{|+\rangle,|-\rangle\right\}$ basis. It establishes that the computational basis is  preferable for the channels with high decoherence rate, while the diagonal basis is the worst choice. The same fact can also be verified from Figure \ref{fig:Kak1} (c). Moreover, the choice of initial state becomes irrelevant in case of AD channels and low decoherence rate PD channels. 

\begin{figure}
\includegraphics[scale=0.6]{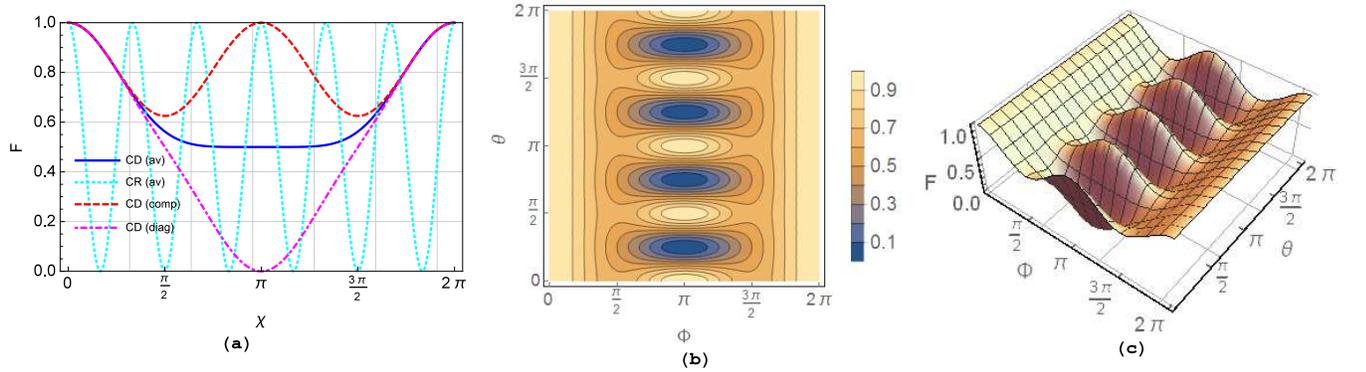}
\caption{\label{fig:Kak2}(Color online) Variation of average fidelity as a function of
noise parameters, when the qubit is subjected to collective noises. In (a), fidelity for specific choice of initial states in the computational $\left\{|0\rangle,|1\rangle\right\}$ and diagonal
$\left\{|+\rangle,|-\rangle\right\}$ basis is shown along with the average fidelity. Here, the noise parameter $\chi=\Theta\,\left(\Phi\right)$ for CR  (CD) noise. In (b) and (c), dependence on the choice of initial state evolving under CD noise is illustrated through a contour and a three-dimensional plots.}
\end{figure}

A similar study on CR noise shows that
fidelity for CR noise is independent of state parameters and is a periodic function of the noise parameter $\Theta$ with period $\frac{\pi}{3}$ (cf. dotted (cyan) line in Fig. \ref{fig:Kak2} (a)). Therefore, there are some specific values of the noise parameter for which the state reaches unaffected. This feature can be attributed to the fact that two arbitrary rotation operators commute with each other. 

On the contrary, the choice of the initial state plays a very important role in the fidelity that can be obtained over the CD noise as shown in Fig. \ref{fig:Kak2}. Specifically, the fidelity over CD channels for the noise parameter $\Phi=\pi$  increases (decreases) to unity (zero) for the choice of computational (diagonal) basis as shown in Fig. \ref{fig:Kak2} (a). Also the average value  of the fidelity, as expected, tends to its lowest value 0.5. The dependence of fidelity on the state parameters is also established using a contour and a three-dimensional plots in Fig. \ref{fig:Kak2} (b) and (c).

One can conclude from the study of analyzing the performance of Kak's protocol in various types of noise models that the commutation, which plays the most important role in the three-stage protocol, between the rotation and noise operators (discussed in the previous subsection) epitomizes the whole scenario.

\section{Conclusion \label{sec:con}}

We have shown that the three-stage QKD protocol proposed by Kak can work as a scheme for  secure direct quantum communication. This provides opportunity to exploit the benefits of single qubit based Kak's protocol in the field of direct communication and their variants as solutions of socioeconomic problems of relevance. However, such a dedicated effort would require serious effort to analyze the feasibility of Kak's protocol under various noise models. 
The present study has established that Kak's protocol would face serious problems in presence of noise. It's further established  that there are certain single qubit states which are preferred over other states for the implementation of Kak's protocol in presence of noise. This is in sharp contrast with the observations made on the basis of the original scheme.  Specifically, in the original protocol, presence of noise was not considered and there was no preference about the states to be chosen to represent bit values 0 and 1. Interestingly,  in the presence of noise such a choice is found to influence the fidelity and thus the performance of the scheme.

It has also been established that Kak-type protocols properly works only under CR noise. It fails under CD (unless decoherence free subspace is used), AD and PD noise models. Logically, a similar study on the effect
of squeezed generalized amplitude damping (SGAD) channel or generalized amplitude
damping channel is also expected to yield similar result (failure). In fact, the same result (failure) is expected over the non-Markovian noise channels \cite{thapliyal2017quantum}. The present work can be extended to include the effect of non-Markovian noise by following the prescription provided in Refs. \cite{thapliyal2017quantum}. However, we have restricted us from doing such an exercise as that would only reveal the same limitation of Kak's protocol.

The protocol can work under the effect of CD noise exploiting decoherence free subspaces for encoding using two-qubit entangled logical qubits instead of single qubits (as discussed in Refs. \cite{boileau2004robust,sharma2016verification} and references therein). However, due to this solution we loose the advantage of single qubit protocols, i.e., a secure protocol without using entanglement. In other words, the use of entangled states would increase the requirement of quantum resources. Further, it may be noted that we have already shown in the recent past that single-qubit-based quantum cryptographic schemes are advantageous when compared to corresponding entangled-state-based counterparts \cite{sharma2016comparative} in the presence of noise. Another possible solution, which would work under any kind of noise model up to moderate decoherence, is to use quantum error correction codes (\cite{pathak2013elements} and references therein). In short, a serious investigation on the error correction scheme specifically designed for Kak's protocol and/or a search for suitable decoherence free subspaces may help Kak's protocol to circumvent the limitations pointed out this paper and thus help its implementation in the realistic situation.

Also in view of our recent results \cite{thapliyal2016quantum}, that the performance of a quantum cryptographic scheme depends upon the complexity of the task in hand and thus rounds of quantum communication involved in accomplishing the task, the three stage protocol is expected to be more affected when compared to a single- or two-stage quantum cryptographic scheme due to multiple rounds of travel of the single qubit through the noisy channel. Thus, despite of its several advantages, Kak's protocol is not preferable in presence of noise.

\textbf{Acknowledgment} KT thanks CSIR, India for the support provided through Senior Research Fellowship. AP thanks Defense Research \& Development Organization (DRDO), India
for the support provided through the project number
ERIP/ER/ 1403163 /M/01/ 1603. 

\bibliographystyle{Final}\bibliography{kakref}

\end{document}